\documentclass[prl,twocolumn,showpacs,preprintnumbers,amsmath,amssymb]{revtex4}

\usepackage{graphicx}

\newcommand{\ket}[1]{\left|#1\right\rangle}
\newcommand{\bra}[1]{\left\langle#1\right|}

\def\*#1{\mathbf{#1}}

\begin{document}

\title{Pure emitter dephasing : a resource for advanced solid-state single photon sources}

\author{Alexia Auff\`eves$^{1}$}%
\author{Jean-Michel G\'erard$^{2}$}%
\author{Jean-Philippe Poizat$^{1}$}%
\affiliation{$^{1}$CEA/CNRS/UJF Joint team " Nanophysics and
semiconductors ",Institut N\'eel-CNRS, BP 166, 25, rue des Martyrs,
38042 Grenoble Cedex 9, France}

\affiliation{$^{2}$CEA/CNRS/UJF Joint team " Nanophysics and
semiconductors", CEA/INAC/SP2M, 17 rue des Martyrs, 38054 Grenoble,
France}

\email{alexia.auffeves@grenoble.cnrs.fr}

\date{\today}

\begin{abstract}
We have computed the spectrum emitted spontaneously by a quantum
dot coupled to an arbitrarily detuned single mode cavity, taking
into account pure dephasing processes. We show that if the emitter
is incoherent, the cavity can efficiently emit photons with its
own spectral characteristics. This effect opens unique
opportunities for the development of devices exploiting both
cavity quantum electrodynamics effects and pure dephasing, such as
wavelength stabilized single photon sources robust against
spectral diffusion.
\end{abstract}

\maketitle

Experiments previously restricted to the field of atomic physics
with isolated two-level atoms and high-Q cavities can nowadays be
implemented using solid-state emitters and optical microcavities,
paving the road towards integrable information processing. The
strong confinement of electron and holes in semiconductor quantum
dots (QDs) results in a discretization of their electronic energy
levels and to an attractive set of atom-like properties, such as
spectrally narrow emission lines at low temperature, and the
ability to generate quantum states of light, like single
photons~\cite{Santori0,Moreau}. At the same time, impressive
progress in the technology of solid-state cavities allow us to
observe cavity quantum electrodynamics effects for a single QD
embedded in a solid-state optical microcavity, like Purcell
effect~\cite{Moreau} and vacuum Rabi splitting
~\cite{Peter,Reithmaier,Yoshie,Hennessy,Press}. Nevertheless, a QD
is far from behaving like an isolated atom. In particular, it
interacts with the phonons of the matrix it is embedded in, giving
rise to sidebands in addition to the so-called zero-phonon line
(ZPL)~\cite{Krummheuer}. At sufficiently low temperature yet, the
emission in the ZPL remains predominant ~\cite{Favero,Besombes},
allowing to model these systems as effective two-level systems.
Another difference is due to the random trapping of carriers in
the vicinity of the QD, leading to the observation of spectral
jumps at long timescale~\cite{LeThomas}, and at shorter
timescales, to the broadening of the homogeneous linewidth of the
transition~\cite{Coolen}. This additional source of decoherence
can be attributed to the lecture by the environment of the state
of the QD, resulting in a loss of indistinguishability of the
emitted photons~\cite{Kiraz}. The best degree of
indistinguishability reported to date for semiconducting QDs,
which has been obtained for single InAs QDs in pillar
microcavities, is of the order of 80$\%$~\cite{Santori, Spiros}.

For this very reason, pure dephasing mechanisms are often
considered as a drawback, likely to reduce severely the potential
interest of QDs for Quantum Information Processing and
Communication (QIPC). In this letter, we show that pure dephasing
can by contrast also be seen as a novel resource for QIPC, that is
specific to solid-state emitters. It has been pointed out recently
that pure emitter dephasing has a crucial influence on the shape
of the emission spectra of QD-cavity
systems~\cite{Cui,Naesby,Noda}, and leads to a strong increase of
the emission at the cavity energy for detuned systems. Moreover,
pure dephasing rate is experimentally controllable by tuning the
temperature of the setup, the pumping rate~\cite{Favero2}, and the
electrical field in the vicinity of the QD~\cite{Berthelot0}. Pure
dephasing provides thus a supplementary degree of freedom,
specific to QDs, which, combined with cavity quantum
electrodynamics effects, offers appealing potentialities to
develop advanced solid-state single photon sources.

\begin{figure}[h,t]c
\begin{center}
\includegraphics[width=8cm]{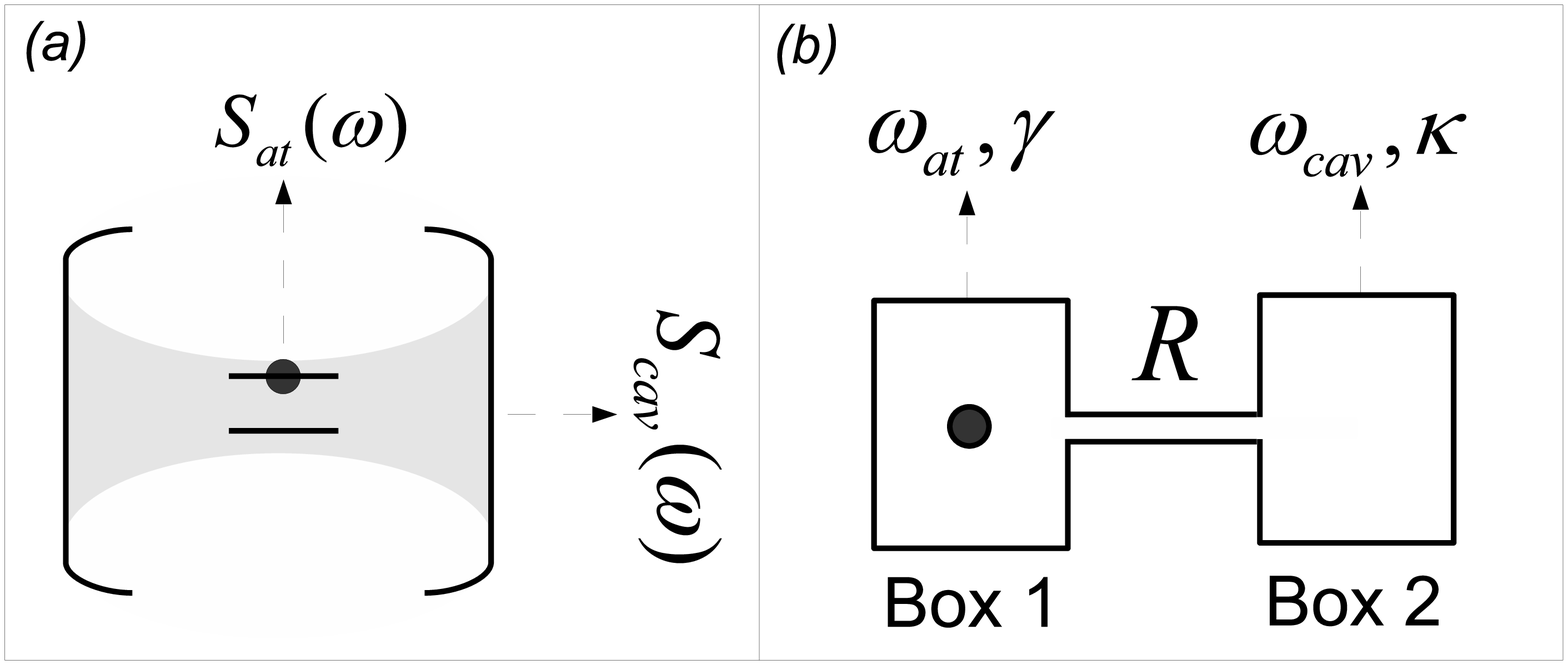}
\caption{\it $(a)$ (Color online) System under study. $(b)$ Equivalent system in
the incoherent emitter regime : two connected boxes exchanging a
quantum of energy.} \label{fig:systeme}
\end{center}
\end{figure}

In this paper, we compute and analyze the spectrum $S_{cav}$
spontaneously emitted by a cavity coupled to a QD initially fed
with a single exciton, and the probability $P_{cav}$ for the
quantum of energy to be emitted in the cavity channel of losses. Like in~\cite{bibi}, the spectrum
$S_{cav}$ is normalized with respect to the frequency $\omega$.
The quantities $S_{cav}$ and $P_{cav}$ can be measured by
a frequency-resolved detector placed in the radiation pattern of
the cavity mode. This generalizes the study held in~\cite{bibi}
where the only considered emitter was an isolated atom, undergoing
no pure dephasing. The system is pictured in
fig.~\ref{fig:systeme}a. The QD's and the cavity's frequencies are
denoted $\omega_0$ and $\omega_{cav}$, the QD-cavity detuning
being $\delta=\omega_{cav}-\omega_0$. The annihilation operator in
the cavity mode is $a$ and the atomic operators
$\sigma_-=\ket{g}\bra{e}$ and $\sigma_+=\ket{e}\bra{g}$, where
$\ket{e}$ and $\ket{g}$ are the upper and ground state of the QD
respectively. In addition to the cavity mode, the QD is coupled to a continuum of leaky modes. The
spontaneous emission rate in this lossy channel is denoted
$\gamma$. This rate is usually measured in time-resolved experiments, the
QD and the cavity being strongly detuned, and ranges from $1$ to
$10~\mu eV$. Pure dephasing is considered in an effective manner
by an additional relaxation term $\gamma^*$ in the evolution
equation of the atomic coherence. If the QD is not coupled to the
cavity mode, the emission lineshape remains Lorentzian, its width
being $\gamma+\gamma^*$. This intrinsic spectrum is further
denoted $S^0_{QD}$. From an experimental point of view, this
modeling is well adapted to the so-called motional narrowing
regime, where Lorentzian emission lines are
observable~\cite{Favero2,Berthelot0,Berthelot}. Pure dephasing
rate in this regime can be increased up to a few hundreds of $\mu
eV$~\cite{Favero2}. On the other hand, the cavity's losses induce
a finite linewidth $\kappa$ of the mode, which typically scales
like $\kappa=100~\mu eV$~\cite{Press,Hennessy}. We denote
$S^0_{cav}$ the empty cavity spectrum as it could be registered in
a transmission experiment. Finally, the QD-cavity coupling
strength is denoted $g$, and checks for the best couplings $g\sim
100~\mu eV$ to $200~\mu eV$~\cite{Hennessy,Peter}. The system is initially prepared
in the state $\ket{e,0}$. Such an initial state can be implemented by quasi-resonantly
pumping the QD as in ref.~\cite{Press,Englund}, so as the QD is properly modeled by a two-level
system.  The evolution of the system is described by a
master equation, and remains restricted in the subspace spanned by the
basis $\{\ket{e,0},\ket{g,1},\ket{g,0}\}$. The populations'
evolution follows the equations

\begin{equation}\label{equ:populations}
\begin{array}{l}
{\displaystyle \frac{d \langle a^\dagger a \rangle }{dt}=-\kappa
\langle a^\dagger a \rangle + g
\langle \sigma_+ a \rangle + g \langle a^\dagger \sigma_- \rangle }\\
{\displaystyle \frac{ d \langle \sigma_+\sigma_- \rangle}
{dt}=-\gamma \langle \sigma_+\sigma_- \rangle - g \langle \sigma_+ a
\rangle - g
\langle a^\dagger \sigma_- \rangle }\\
{\displaystyle \frac{ d \langle \sigma_+a \rangle} {dt}=-i\delta
\langle \sigma_+a \rangle -\frac{\gamma+\gamma^*+\kappa}{2} \langle
\sigma_+a \rangle +g (\langle \sigma_+\sigma_- \rangle - \langle
a^\dagger a \rangle}).
\end{array}
\end{equation}

We have computed the spectrum $S_{cav}$ emitted by the
cavity using Glauber's formula~\cite{Glauber}, it fulfills

\begin{equation}\label{Scav}
S_{cav} \propto {\displaystyle
\frac{1}{|\omega-\lambda_+|^2}\frac{1}{|\omega-\lambda_-|^2}}
\end{equation}

where $\lambda_+$ and $\lambda_-$ are the roots of the secular
equation

\begin{equation}\label{roots}
(\tilde{\omega}_{at}-\omega)(\tilde{\omega}_{cav}-\omega)-g^2=0.
\end{equation}

\begin{figure}[h,t]c
\begin{center}
\includegraphics[width=10cm]{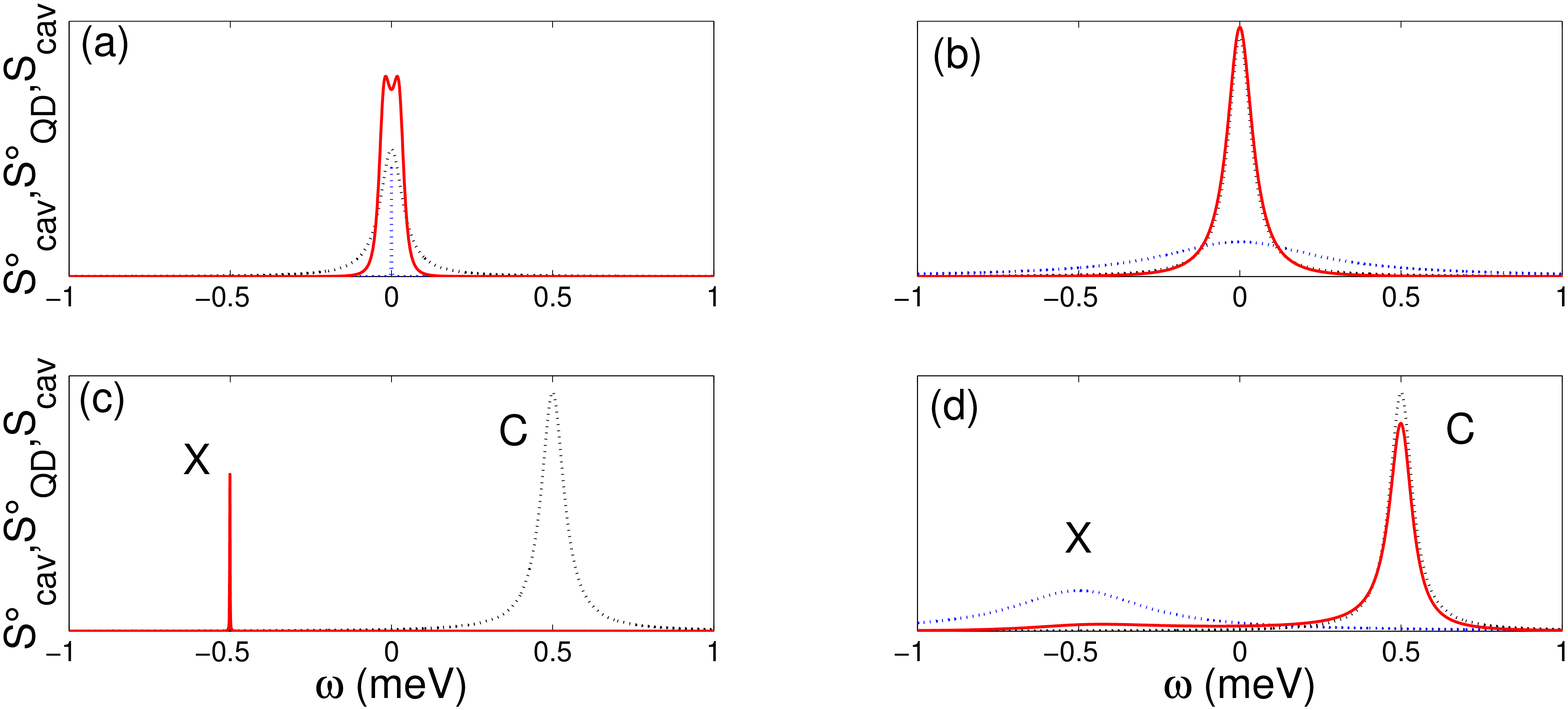}
\caption{\it (Color online) Spectrum $S_{cav}$ emitted by the coupled cavity (red
solid line) and spectra emitted by the QD $S_{QD}^0$ (blue dotted
line) and by the cavity $S_{cav}^0$ (black dashed line) if they
were uncoupled, for $g=35~\mu eV, \kappa=85~\mu eV, \gamma=1~\mu
eV$, in the resonant case ((a) and (b)), and in the case $\delta=1
meV$ ((c) and (d)). (a) and (c) : $\gamma^*=0$, (b) and (d) :
$\gamma^*=0.5 meV$. } \label{fig:spec}
\end{center}
\end{figure}

We have introduced the complex frequencies of the QD and of the
cavity in the absence of coupling
$\tilde{\omega}_{0}=\omega_0-i\gamma/2$,
$\tilde{\omega}_{cav}=\omega_{cav}-i\kappa/2$. If the coupling is
weak with respect to the detuning or to the QD's and cavity
widths, the roots $\lambda_+$ and $\lambda_-$ equal the intrinsic
frequencies $\tilde{\omega}_{0}$ and $\tilde{\omega}_{cav}$, and
the spectrum emitted by the cavity simply writes

\begin{equation}\label{equ:filtre}
S_{cav}\propto S^0_{cav}S^0_{QD}, \end{equation}

showing that the cavity behaves as a spectral filter for the light
emitted by the QD. As a consequence, if the QD's emission line is
narrower than the cavity linewidth ($\gamma+\gamma^* < \kappa$),
photons are mostly emitted at the QD's frequency, and at the
cavity's frequency in the opposite case as it appears in
fig.~\ref{fig:spec}. We took the parameters of ref.~\cite{Press},
and considered the evolution of the spectrum emitted by the cavity
when, starting from $\delta=0$ (fig.~\ref{fig:spec}a and
~\ref{fig:spec}b), one switches to $\delta=~1~meV\sim 11\kappa$
(fig.~\ref{fig:spec}c and ~\ref{fig:spec}d). This quantity is the typical
amplitude of spectral jumps as they can be observed
for a single nanocrystal coupled to a microsphere~\cite{LeThomas}. 
It also provides an upper 
bound for the range of spectral diffusion affecting the emission
line of self-assembled QDs~\cite{Berthelot,Favero}. First we consider
the case when $\gamma^*\sim0$ which corresponds to the
experimental conditions of ref~\cite{Press}. At resonance one 
can almost observe the vacuum Rabi doublet (fig.~\ref{fig:spec}a). If the 
excitonic transition's frequency
changes, the cavity emits photons at this new
frequency (fig.~\ref{fig:spec}c). On the contrary, if
$\gamma^*$ is increased so that it overcomes the cavity linewidth
$\kappa$ (we took $\gamma^*=0.5 meV$ which corresponds to
reasonable experimental parameters), photons are emitted at the
cavity frequency (fig.~\ref{fig:spec}d). We stress that in the
regime of parameters we consider, the spectrum $S_{cav}$ of the
photons emitted by the detuned cavity exactly matches the spectrum
$S_{cav}^0$ of the empty cavity mode. This is because the QD's
spectrum $S^0_{QD}$ is flat in the vicinity of the cavity's
frequency (eq.~\ref{equ:filtre}) : the QD behaves as an internal
white light source. Interestingly, increasing the pure dephasing
rate provides a safe way to measure the intrinsic quality factor
of a solid-state cavity, even though it is only fed with a single
emitter. This mechanism may also allow to
enforce a single photon source to emit photons at the cavity's
frequency, making it robust against spectral jumps and spectral diffusion. We emphasize
that the figures are plotted with realistic experimental
parameters, showing that this functionality can be achieved with
state of the art QDs and cavities, provided one manages to control
the environment of the QD.

Until now we have only paid attention to the normalized spectrum $S_{cav}$
emitted by the cavity mode. In the framework of photonic devices it is also essential to estimate
the efficiency $P_{cav}$ of the emission process. First note that
if pure dephasing is sufficiently strong, coherences can be
adiabatically eliminated in the set of
equations~\ref{equ:populations}. The dynamics of the system is now
described by classical probabilities and corresponds to the
diffusion of a particle between an atomic box and a cavity box,
each box having a probability $\gamma$ (resp. $\kappa$) per unit
of time to lose the particle as it is pictured in
figure~\ref{fig:systeme}b. The coupling rate $R$ between the two
boxes fulfills

\begin{equation}\label{equ:R}
R(\gamma^*,\delta)=\frac{4g^2}{\kappa+\gamma+\gamma^*}\frac{1}{1+\left({\displaystyle\frac{2\delta}{\kappa+\gamma+\gamma^*}}\right)^2},
\end{equation}

and evolves like the overlap between the uncoupled QD and cavity
spectra $S_{QD}^0$ and $S_{cav}^0$. Using this picture the
efficiency $P_{cav}$ can easily be computed, one finds

\begin{equation}\label{equ:Pcav}
P_{cav}=\frac{\kappa}{\kappa+\gamma}\frac{{\cal C}}{1+{\cal C}},
\end{equation}

where we have introduced the cooperativity ${\cal C}$, fulfilling

\begin{equation}\label{equ:coop}
{\cal C}=R\left(\frac{1}{\kappa}+\frac{1}{\gamma}\right).
\end{equation}

Note that although these expressions are most easily derived in
the incoherent regime, they are in fact valid for all values of
the pure dephasing rate $\gamma^*$. In particular, if
$\gamma^*=0$, one recovers the form computed in ref~\cite{Cui} in
the resonant case and in the absence of pure dephasing.

\begin{figure}[h,t]c
\begin{center}
\includegraphics[width=10cm]{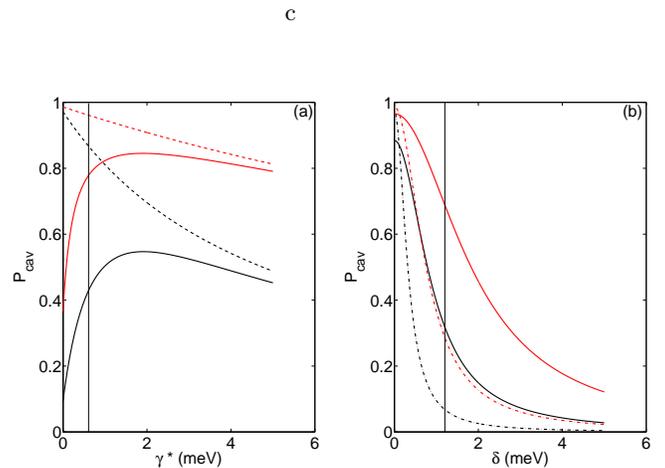}
\caption{\it (Color online) Efficiency $P_{cav}$ as
a function of (a) the pure dephasing rate $\gamma^*$ (b) the QD-cavity
detuning $\delta$. Black plots :  $g=35~\mu$eV, $\kappa=85~\mu$eV, 
$\gamma=1~\mu$eV (parameters of ref.~\cite{Press} 
and fig.~\ref{fig:spec}). Red plots : $g=76~\mu$eV, $\kappa=100~\mu$eV, 
$\gamma=1~\mu$eV (parameters of ref.~\cite{Hennessy}). (a) 
Dashed line : $\delta=0$. Solid line : $\delta=1~$meV. (b) Dash-dotted line : 
$\gamma^*=0$. Solid line : $\gamma^*=0.5$~meV.} \label{fig:eff}
\end{center}
\end{figure}

We have represented in fig.~\ref{fig:eff}a the evolution of the
efficiency $P_{cav}$  of the cavity emission process as a function of the
pure dephasing rate, in the resonant case and in the case where
the QD and the cavity are detuned by $\delta=1~$meV. We considered
the parameters of ref.~\cite{Press} and ref~\cite{Hennessy}. In
the resonant case, the efficiency drops because the
overlap between the uncoupled spectra $S^0_{QD}$ and $S^0_{cav}$
decreases, and thus the cooperativity ${\cal C}$ of the
corresponding single-photon source as it is mentioned above. On
the contrary, when the QD and the cavity are detuned, the overlap
starts to increase with respect to the pure dephasing rate, so as
the efficiency of the process as it appears in the figure. 
The fact that pure dephasing enhances the effective coupling between
a QD and a detuned cavity mode may provide a partial explanation for the lasing 
observed for a photonic crystal cavity mode coupled to a few detuned quantum dots ~\cite{Strauf}, 
as it was suggested in~\cite{Naesby}.
The optimum is reached for $\kappa+\gamma^*=2\delta$, and the maximal
cooperativity equals $g^2/\gamma \delta$. The enhancement that could be reached in the
case of ref~\cite{Hennessy} is more spectacular, as the coupling
strength $g$ is bigger than in ref~\cite{Press}. 

The evolution of the efficiency $P_{cav}$  as a function of the detuning 
$\delta$ is plotted in fig~\ref{fig:eff}b, for $\gamma^*=0$ and $\gamma^*=0.5$~meV.
It appears that if $\delta \leq \kappa$, a single photon source undergoing pure dephasing 
is more efficient than the same source perfectly isolated from its environment.
This effect could be exploited to increase the
fabrication yield of efficient single-photon sources prepared with a sample
of inhomogeneously broadened QDs. Another advantage of pure
dephasing is that it makes the efficiency of a single-photon source less
sensitive to spectral jumps or spectral diffusion. 
As an example, one consideres how the efficiency of
a single photon source is affected if the QD and the cavity,
initially on resonance, are detuned by $\delta=1~$meV. 
For the parameters of ref.~\cite{Press},
without pure dephasing, the efficiency of the device 
drops from $97~\%$ to $10~\%$, whereas it
only jumps from $90~\%$ to $40~\%$ in the presence of pure
decoherence. With the parameters of ref.~\cite{Hennessy}, the
efficiency drops from $99~\%$ to $37~\%$ without pure dephasing,
whereas with pure dephasing, it only drops from $97~\%$ to
$76~\%$. As a consequence, pure dephasing not only allows to
stabilize the frequency of the device, but also its
efficiency.

At this stage, one can highlight a major difference between a QD
and an isolated atom. Cavity filtering requires at least the
condition $\gamma+\gamma^*
> \kappa$. On the other hand, a necessary condition to have 
efficient cavity emission is $\kappa > \gamma$. It is obvious that
the two conditions cannot match if $\gamma^*=0$ : as a consequence,
there may be no efficient cavity filtering for an isolated atom.
Pure dephasing appears as a supplementary degree of freedom,
specific to QDs, allowing the engineering of their emission linewidth and
their losses in a decoupled way. As a consequence, it is
possible to combine low atomic losses and cavity filtering to
achieve an efficient wavelength stabilized single photon source.
For example, in the case of ref.~\cite{Hennessy}, with $\delta=1$~meV, the filtering condition
is fulfilled with $\gamma^*=0.5$~meV, and the efficiency of the process 
is $76~\%$ as it was mentioned above.  
Moreover, the process is all the more efficient than the QD's losses are reduced;
namely, very high values for $P_{cav}$ can be reached, provided
 $\gamma\ll \kappa,R(\gamma^*,\delta)$(see eq.~\ref{equ:coop}).
This is intuitive, as the quantum has no
other option apart from being released in the cavity channel of
losses :  the ideal single photon source is
nothing but a shielded atomic box connected to a lossy cavity. 
Considering again the case of ref.~\cite{Hennessy}, the local density of leaky
modes could be engineered to reduce the corresponding sponteanous emission 
rate to $\gamma=0.1\mu$eV. In this case, the efficiency reaches $96~\%$.
With these parameters, most photons are spontaneously emitted at the cavity frequency, 
even though the cavity is detuned from the QD, providing efficient energy conversion.

To conclude, it is nice to notice that the emitter's decoherence
can also be exploited to develop a source of indistinguishable
photons. Indistinguishable photons are resources in the frame of
quantum computation with linear optics~\cite{KLM01}. A necessary
condition of indistinguishability is that photons are
Fourier-transform limited, namely that the spectral width $\delta
\nu$ of the photonic peak corresponds to the inverse of its
duration $\tau$. Like it appears in eq.~\ref{Scav}, a cavity
coupled to a QD emits photons whose frequency and linewidth are
imposed by the narrowest root of eq.~\ref{roots}. In the usual
approach, pure dephasing is very weak, and the linewidth of the
emitted photons corresponds to the linewidth of the QD "dressed"
by the cavity, namely in the resonant case, $\delta
\nu=\gamma+\gamma^*+4g^2/\kappa$. The duration $\tau$ of the
wavepacket checks $\tau^{-1}=\gamma+\kappa / 4g^2$, which
corresponds to the relaxation of the QD in the cavity in the
Purcell regime. To recover indistinguishability, the usual strategy
consists in reducing $\gamma^*$ by lowering the temperature and
quasi-resonantly pumping the system, and by increasing the
relaxation rate by Purcell effect~\cite{Santori,Spiros}.
Nevertheless, this strategy has its limits as it appears in
ref~\cite{Kiraz}. In particular, the relaxation time must remain
large enough so that the process remains insensitive to the jitter
due to the optical pumping time. Moreover, as the spectral
characteristics of the emitted photons are governed by the
emitter, one is linearly sensitive to any variation in the
frequency or in the emission linewidth of the QD.

On the contrary, an alternative strategy consists in increasing
the pure dephasing rate, so that it overcomes the cavity
linewidth. Provided the QD's spectrum is sufficiently flat in the
vicinity of the cavity's frequency (condition of white light
regime), the frequency and the width of the emitted photons are
imposed by the cavity as it was explained above. Moreover, it can
easily be shown that the lifetime $\tau$ of the quantum in the
atom-cavity system checks

\begin{equation}
\tau^{-1}=\frac{\kappa+\gamma}{2}+R-\sqrt{\left(\frac{\kappa-\gamma}{2}\right)^2+R^2}
\end{equation}

If $\gamma=\kappa$, this lifetime equals the cavity lifetime, and
the emitted photons are Fourier-transform limited, and thus
indistinguishable. In this approach, the cavity is used as an
integrated spectral filter that restores the temporal
indistinguishability of the emitted photons, initially degraded by
the interaction of the QD with its environment~\cite{Kiraz}. It
may provide solid-state physicists with an original method to
generate photons showing a degree of indistinguishability as high
as desired, and relaxes the constraint on low temperature.
Nevertheless, this method is intrinsically limited. The white
light regime is reached indeed when there is a poor overlap
between the QD's and cavity's uncoupled spectra, and as a
consequence, a poor cooperativity ${\cal C}$. Thus there is a
tradeoff between the degree of indistinguishability of the emitted
photons and the efficiency of the source, which is inherent to any
method based on spectral filtering. To have a glimpse of the
performances of the device, we have considered the wavelength
stabilized single photon source studied above, supposing now that
$\kappa=\gamma=10~\mu eV$. In practice, the fabrication of such a
device should be doable in the near future, as nowadays, state of
the art QD and cavities already allow to achieve $\kappa = 10
\gamma$~\cite{Reithmaier,Yoshie,Peter,Hennessy,Press}, and
impressive progress has been recently witnessed for semiconductor
based cavities such as micropillars~\cite{Reitzenstein},
microdisks~\cite{Srivanasan} or photonic crystal
cavities~\cite{Weidner}. First we have computed the degree of
indistinguishability $d=\tau / \delta\nu$ of the emitted photon
in the resonant case, which checks $d=80\%$. A convenient way to
reach the white light regime is to strongly detune the QD from the
cavity mode. For a detuning $\delta=1.5~meV=150\kappa$, the
efficiency of the source is $3\%$, whereas a degree of
indistinguishability $d=97\%$ is reached. The tradeoff between
efficiency and indistinguishability clearly appears here, as this
method only consists in spectral filtering "on chip".
Nevertheless, it paves the road towards the generation of
indistinguishable photons at finite temperature.

As it is underlined in ref.~\cite{Naesby,Noda}, pure
dephasing could explain part of the results obtained for
the emission of a semi-conducting cavity coupled to a detuned
QD~\cite{Press,Hennessy}. Providing a clear understanding of these 
striking properties is very challenging. All explanations, including ours,
invoke the broadening of the QD's emission line, combined with cavity filtering.
Within our model, the QD remains properly described by a two-level system.
In other explanations, the broadening of the QD's emission line is due to 
a continuum of final states for the excitonic transition : namely, Press et al \cite{Press}
invoque phonon-assisted processes, and Kaniber et al~\cite{Kaniber}, a continuum of
final charged states of the dot. The next step will be
to take into account the cavity, to quantitatively estimate the efficiency 
of each relaxation channel. 
Note that within the frame of these three models, the correlation measurements
achieved on the cavity mode should be
antibunched, which is not the case in ref.~\cite{Hennessy}. As it is underlined 
in~\cite{Noda}, the QD is non-resonantly pumped in this reference, which may 
induce multi-photonic processes. As it is stated in the beginning, our model
is better adapted to the experiments where the QD is properly described
by a two-level atom, which is the case if it is quasi-resonantly pumped
like in~\cite{Press}. Let's stress that resonant  excitation like in~\cite{Englund}
seems to further reduce multi-photon processes. This excitation regime, that avoids jitter and spectral
fluctuation, is quite well described by the model exposed in this paper.

We have shown that pure decoherence is a degree of freedom specific
to solid-state emitters, giving rise to unexpected regimes for
cavity quantum electrodynamics. Provided one succeeds in
controlling their environment, quantum dots could in the near
future be used as toy models to explore these new regimes, making
them free from the two-level atom paradigm. It also appears that
pure dephasing, far from being a drawback, is a resource that can
be exploited to develop advanced nano-photonic devices like
frequency stabilized solid-state single-photon sources, opening
promising perspective for quantum computation on chip.

\section{Acknowledgments}

Alexia Auff\`eves thanks Marcelo Fran\c{c}a Santos, Xavier
Letartre, Pierre Viktorovitch and Christian Seassal  for all the
conversations. The authors thank J. Moerk for pointing to our
attention the crucial role of emitter dephasing on QD-cavity
spectra prior to the publication of~\cite{Naesby}. Jean-Michel
G\'erard acknowledges partial support from the IST-FET QPhoton
project. This work was supported by IP European project 'QAP'
(contract number 15848).

\end{document}